\documentclass[slac_one,amsmath,nofootinbib]{revtex4}
\usepackage{graphicx,axodraw}
\usepackage{xspace}
\usepackage{fancyhdr}
\newcommand{\beq}{\begin{equation}}
\newcommand{\eeq}{\end{equation}}
\newcommand{\beqa}{\begin{eqnarray}}
\newcommand{\eeqa}{\end{eqnarray}}

\newcommand{\ee}{e^+e^-}

\newcommand{\caesar}{\textsc{caesar}\xspace}

\begin{document}

\title{
Resummation of event-shapes at hadron-hadron colliders
}

\author{G.~Zanderighi} \affiliation{CERN, TH Division, CH-1211, Geneva
  23, Switzerland\\
 Fermilab, P.O. Box 500, 60510 Batavia, IL, US}
\begin{abstract}
  We point out that a study of event shapes at hadron colliders allows
  to explore novel aspects of QCD. These studies are today made easier
  by the development of a program which automates the resummation.
\end{abstract}
\maketitle
\thispagestyle{fancy}
\section{introduction}
In the last decades $\ee$ and subsequently DIS event shapes have
proved to be a very powerful tool to study the all order properties of
QCD (for a review see e.\ g.~\cite{DasSalReview}). They allowed
measurements of $\alpha_s$ and its renormalization group running,
cross checks of the colour factors, tuning of Monte Carlos and studies
of analytical hadronization models (e.\ g.\ the renormalon inspired
power correction approach~\cite{Dokshitzer:1995qm}).
A key ingredient, which made those studies possible, was the fact that
the coherence of QCD radiation allows one to reorganise the
perturbative (PT) expansion so as to resum the leading and
next-to-leading contributions. This is needed in the description of
the more exclusive phase space region where the event-shape's value is
very small. Here the real-virtual cancellation works only partially,
leaving behind large logarithms of the value of the event shape, which
need to be resummed to all orders. This machinery is today
well-understood. 

Despite their great success in $e^-e^+$ and in DIS, event shapes have
been largely neglected at hadron-hadron colliders (hhc), the only
measurements, to our knowledge, being a measurement of the broadening
by CDF in 1991~\cite{CDF-Broadening} and a measurement of the thrust
by D0 in 2002~\cite{D0Thrust}.
One might consider the study of event-shapes at hadron colliders of
limited interest, being almost a repetition of what has already been
studied in great detail in $e^+e^-$ and in DIS. The main point of this
talk is to state that this is by no means the case.

First, it is important to mention that till date, comparison of
next-to-leading logarithmic (NLL) resummed predictions with data have
been carried out {\em only} for observables which vanish in the 2-jet
limit: 2-jet $\ee$ event shape variables, [1+1] event-shapes in DIS
and Drell-Yan vector boson production.
At hhc, where two jets are already present in the initial state, any
study of the final state goes beyond the known and well-tested domain.
Dijet production in hhc represents a field where both, resummation and
power corrections are completely untested.

It is however clear that the experimental environment in hhc is much
more difficult.  Due to the presence of the beam it is less clean then
in $\ee$ and does not have the benefit of DIS experiments where one
can considerably vary the hard scale ($Q$) of the process, thereby
allowing a single experiment to explore a whole range of widely
different scales.
The study of a single or just few event shapes is then not enough to
balance the experimentally more challenging environment. In order to
extract some information from hhc experiments one needs to study a
whole range of event shapes.

Following an old-fashioned, analytic approach, this would be barely
possible: the complexity of analytical resummations increases
drastically in multi-jet event shapes, e.\ g.\ the 3-jet thrust-minor
calculation in the ``simple'' $\ee$ environment requires already the
evaluation of five Mellin integrals~\cite{Banfi:2000si}.  In the study
of hhc dijet event-shapes one has four QCD emitters and, additionally,
one has to face new complications arising from the fact that the
colour structure is not diagonal~\cite{ColEv}. Even if possible, one
must admit that facing similar, tedious calculations for a whole range
of hhc dijet event shapes does not sound like the most appealing
programme.
However, recently an automated way of resuming event-shapes in $\ee$,
DIS and hhc was developed and implemented in the computer program
\caesar ~\cite{BSZauto}.
This made it possible to resum a whole range of observables in hhc and
opened up the possibility to new phenomenological studies at the
Tevatron, some of which are currently under way.

\section{Hadron-hadron collider event shapes} 
An essential point when performing an automated resummation is that
current understanding allows us to resum only observables which enjoy
certain properties.
A goal of an automated resummation tool is that it can be used without
a detailed understanding of the analytical properties of the
observable under consideration. It is therefore an essential property
of \caesar that it can test whether an observable is in its scope and
only in this case it performs the resummation.

The detailed list of properties the observable has to fulfill has been
given in~\cite{BSZauto}. Most of these properties were satisfied by
all observables resummed analytically in the previous years and are
insofar not very limiting.
The experimentally more limiting requirement is that of {\em
  continuous globalness}.  An observable is defined to be {\em
  global}~\cite{Dasgupta:2001sh} if it is sensitive to emissions
everywhere in phase space (and is deemed to be {\em non-global}
otherwise).  Typical examples of non-global event-shapes are
single-hemisphere observables in $\ee$ and many current hemisphere DIS
event shapes.  An important lesson which was learned from DIS is that
{\em not measuring everywhere} does not necessarily mean {\em
  non-global}.  There are event-shapes in DIS which measure particles
only in the current hemisphere but are nonetheless sensitive to
emissions in the remnant hemisphere, typically through the recoil of a
hard parton.  Often the sharp distinction between global and
non-global event shapes is too coarse and it is useful to distinguish
between various degrees of (non-)globalness.  A more restricted class
of global observables is that of {\em continuous global} ones, those
for which the dependence on the transverse momentum is independent of
the emission's direction (i.\ e.\ the observable should scale
according to the same power of $k_t$
everywhere)~\cite{Dasgupta:2001sh}.  \caesar is limited to those
observables.

Having established the important criterion of continuous globalness,
the first question is if there are hhc observables which do satisfy
it. Detectors can cover always only a limited rapidity range (at the
Tevatron $\eta_{\rm max} \sim 3.5$, at the LHC $\eta_{\rm max} \sim
5$).  This makes observables non-global, since there might be
emissions at $\eta > \eta_{\rm max}$ which emit radiation in the
observed region. NLL resummations do not account for such effects.
The theoretical requirement of globalness seems therefore to conflict
with the experimental possibility of current experiments, thus making
the resummation of global hhc event shapes a nice theoretical
exercise, but without much practical use.
Luckily, this is not the case.  We present here three classes of
observables which are specifically designed to solve the tension
between experimental and theoretical needs~\cite{hhres}
\begin{enumerate}
\item \underline{Directly global event shapes:} 
  take $\eta_{\rm max}$ as large as experimentally possible and extend
  usual $\ee$ event-shapes by simply formulating them in the
  transverse plane. Use then the fact that emissions contribute
  significantly only if $v \simeq e^{-(a+b_\ell)\eta_{\rm
      max}}$.\footnote{One of the assumptions of \caesar is that the
    event shape has the following functional behaviour $V(\tilde p, k)
    = d_{\ell}\left(\frac{k_t}{Q}\right)^a e^{-b_\ell \eta }
    g_{\ell}(\phi)$ when a single, soft emission is emitted collinear
    to the hard leg $\ell$. Here $k_t$, $\eta$ and $\phi$ are measured
    wrt the leg $\ell$. } This implies that standard NLL resummations
  are still valid, but in a limited region where $v$ is not too small
  ($v \le e^{-(a+b_\ell)\eta_{\rm max}}$).  These observables
  typically have $b_\ell=0$ for incoming legs.
  
\item \underline{ Indirectly global event shapes with recoil term:}
  Define a central region ${\cal C}$ well away from the forward
  detector edges and define the observable in terms of only these
  emissions. Add then a recoil term which is constructed only with
  emissions in ${\cal C}$, but which is non zero if there are
  emissions in ${\cal \bar C}$ (e.\ g.\ the vectorial sum of all
  transverse momenta in ${\cal \bar C}$, or any power of it, to ensure
  continuous-globalness). Due to the presence of the recoil term
  exponentiation to NLL holds only in impact-parameter space, this is
  due to the fact the smallness of the observable is not necessarily
  to be attributed to a Sudakov effect, but could be due to a
  vectorial cancellation between harder emissions. This again limits
  the range of applicability of resummed calculations.
\item \underline{Directly global event shapes with exponentially
    suppressed forward terms:}
  Define a central region ${\cal C}$ and define the observable in
  terms of only these emissions. Add then a term which is constructed
  with emissions in ${\cal \bar C}$ but which is not sensitive to
  details of the emission pattern, i.\ e.\ the term is exponentially
  suppressed in the rapidity of the emissions. The presence of the
  exponentially suppressed term reduces the sensitivity to the region
  ${\cal \bar C}$, but guarantees that the observable is global.
  These observables typically have $b_\ell=a$ for incoming legs.
\end{enumerate}
By construction these observables reconcile the experimental and
theoretical needs, making them a solid basis for phenomenological
studies. Additionally they have complementary properties e.\ g.\ with
respect to the sensitivity to the beam, making them a rich source of
information.
It is then possible to not only consider extensions of typical $\ee$
and DIS event-shapes (thrust, broadening, jet-masses, \dots) but also
to design new observables which arbitrarily enhance or suppress the
importance of the beam radiation.  \footnote{Despite this, in hhc
  studies of event-shapes one alwasy has to deal with an omnipresent
  underlying event. It is therefore important that these studies be
  supplemented by similar studies of multi-jet event shapes both in
  $\ee$ (3-jet) and in DIS [1+2], for which NLL predictions and power
  corrections exist~\cite{eeDIS3jet}.}

To conclude, there are several aspects of QCD which can be studied by
analyzing hhc dijet event-shapes:
\begin{itemize}
\vspace{-.15cm}
\item resummation for dijet event-shapes accounts for multiple soft,
  large angle radiation emitted from a four-parton system. Here
  quantum evolution of colour~\cite{ColEv} enters the game.  These
  novel perturbative QCD colour evolution structures have never been
  investigated before; ~\footnote{Notice that such effects enter also
    top threshold corrections.}  \vspace{-.15cm}
\item studies of hadronization corrections and power-corrections in
  multi-jet events;  
\vspace{-.15cm}
\item studies of underlying event, these are made easier by the fact
  that the forward sensitivity (to beam-fragmentation) can be
  arbitrary tuned.  \vspace{-.15cm}
\end{itemize}

\section{Conclusions}
Automated resummation makes it possible to study a large number of
event shapes at hadron-hadron colliders. These observables can be
designed so as to have complementary properties, e.g. the sensitivity
to beam-fragmentation can be arbitrarily tuned.
Studies of event-shapes at the Tevatron are currently under way. We
very much hope that these measurements will turn out to be as
successful as those in $e^+e^-$ and DIS collider experiments.

\begin{acknowledgments}
  I thank A. Banfi and G.P. Salam for collaboration in this project.
\end{acknowledgments}

\end{document}